\documentclass[conference]{IEEEtran}
\usepackage{amsmath,amsfonts}
\usepackage{algorithmic}
\usepackage{array}
\usepackage{subfigure}
\usepackage{textcomp}
\usepackage{stfloats}
\usepackage{url}
\usepackage{verbatim}
\usepackage{graphicx}
\usepackage{cite}
\usepackage{booktabs} 
\usepackage{mathrsfs}
\usepackage{amssymb}
\usepackage{multirow}
\usepackage[ruled]{algorithm2e}
\usepackage{float}
\usepackage{xcolor}
\usepackage[colorlinks=true,linkcolor=black,citecolor=blue,urlcolor=blue,]{hyperref}
\begin{document}

\title{Cross-Modal Semantic Communication for Heterogeneous Collaborative Perception}

\author{\IEEEauthorblockN{Mingyi Lu$^{1}$, Guowei Liu$^{1}$, Le Liang$^{1,2}$, Chongtao Guo$^{3}$, Hao Ye$^{4}$, Shi Jin$^{1}$}
\IEEEauthorblockA{$^1$School of Information Science and Engineering, Southeast University, Nanjing 210096, China}
\IEEEauthorblockA{$^2$Purple Mountain Laboratories, Nanjing 211111, China}
\IEEEauthorblockA{$^3$College of Electronics and Information
     Engineering, Shenzhen University, Shenzhen 518060, China}
\IEEEauthorblockA{$^4$ Department of Electrical and Computer Engineering, University of California, Santa Cruz, CA 95064, USA}

Emails: \{lu\_my, grownliu, lliang, jinshi\}@seu.edu.cn, ctguo@szu.edu.cn, yehao@ucsc.edu.
}

\maketitle

\begin{abstract}
  
Collaborative perception, an emerging paradigm in autonomous driving, has been introduced to mitigate the limitations of single-vehicle systems, such as limited sensor range and occlusion. To improve the robustness of inter-vehicle data sharing, semantic communication has recently further been integrated into collaborative perception systems to enhance overall performance. However, practical deployment of such systems is challenged by the heterogeneity of sensors across different connected autonomous vehicles (CAVs). This diversity in perceptual data complicates the design of a unified communication framework and impedes the effective fusion of shared information. To address this challenge, we propose a novel cross-modal semantic communication (CMSC) framework to facilitate effective collaboration among CAVs with disparate sensor configurations. Specifically, the framework first transforms heterogeneous perceptual features from different sensor modalities into a unified and standardized semantic space. Subsequently, encoding, transmission, and decoding are performed within this semantic space, enabling seamless and effective information fusion. Extensive experiments demonstrate that CMSC achieves significantly stronger perception performance than existing methods, particularly in low signal-to-noise ratio (SNR) regimes.

\end{abstract}


\section{Introduction}	

\IEEEPARstart{W}{ith} the rapid advancement of artificial intelligence, autonomous driving has emerged as one of its most significant and transformative application domains. As a crucial component of autonomous vehicles, the perception module is responsible for understanding the surrounding environment and supplying essential inputs to downstream planning and control. However, single-vehicle perception is fundamentally constrained by sensor range and occlusions. To overcome these limitations, collaborative perception has been proposed, enabling connected autonomous vehicles (CAVs) to form a more comprehensive and accurate environmental understanding by sharing their perceptual information.

Based on the stage at which perception information is shared, collaborative perception can be categorized into three primary schemes: early, intermediate, and late collaboration \cite{review8}. Early collaboration involves the direct exchange of raw sensor data, providing the most comprehensive information at the cost of substantial communication overhead. Conversely, late collaboration shares only the final perception output, significantly reducing bandwidth requirements but suffering from severe loss of contextual information. Striking a balance between these two extremes, intermediate collaboration has emerged as the predominant framework. By sharing and fusing extracted perceptual features, this approach effectively reconciles perceptual performance with communication efficiency~\cite{v2vnet,where2comm,lyd}.


\begin{figure}[t!]
  \centering
  \includegraphics[width=0.8\columnwidth, height=0.48\columnwidth]{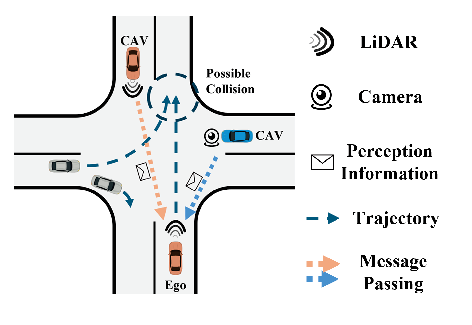}
  \caption{An illustration of heterogeneous collaboration perception.}
  \label{scenarios}
\end{figure}

Most existing studies on intermediate collaboration implicitly assume ideal wireless communication channels. However, in practical vehicular networks, factors such as channel fading and noise inevitably corrupt transmitted data, thereby degrading the performance of collaborative perception. 
To address this challenge, semantic communication based on joint source–channel coding (JSCC) has emerged as a promising solution~\cite{ai}. By extracting and transmitting task-oriented semantic features, this paradigm enhances both bandwidth efficiency and robustness against channel impairments. Initial works in this domain demonstrated the superiority of JSCC-based frameworks for collaborative perception over conventional separate source and channel coding schemes \cite{franklin,harq}. 

Despite these advances, the aforementioned semantic communication frameworks introduced a simplifying assumption that all collaborating vehicles possess homogeneous sensor configurations. This assumption renders most existing systems inherently single-modal and incapable of resolving semantic inconsistencies caused by heterogeneous sensor types. However, vehicles may be equipped with different types of sensors, leading to inherent heterogeneity in perceptual information. Fig.~\ref{scenarios} illustrates such a scenario, depicting collaborative perception between an ego vehicle and CAVs equipped with diverse sensor modalities. While some prior studies have attempted to address the heterogeneity problem in collaborative perception~\cite{heal,stamp}, its reliance on ideal channel assumptions renders it impractical for real-world deployment. Consequently, designing a semantic communication framework to achieve efficient and robust heterogeneous collaborative perception remains a key challenge.

To address this challenge, this paper introduces a novel cross-modal semantic communication (CMSC) framework for heterogeneous collaborative perception. Specifically, we propose a semantic converter that transforms heterogeneous features from disparate sensor modalities into a unified, modality-invariant semantic space to achieve modal alignment. Furthermore, to enhance transmission efficiency, we integrate the framework with a semantic importance-aware strategy that intelligently prioritizes and transmits the most task-relevant information from this unified semantic space. To the best of our knowledge, this work represents the first multi-modal end-to-end semantic communication solution for heterogeneous collaborative perception, enabling robust and efficient collaboration among CAVs with disparate sensor modalities.

\section{System Model}
\begin{figure*}[htbp]
  \centering
  \includegraphics[width=0.75\linewidth]{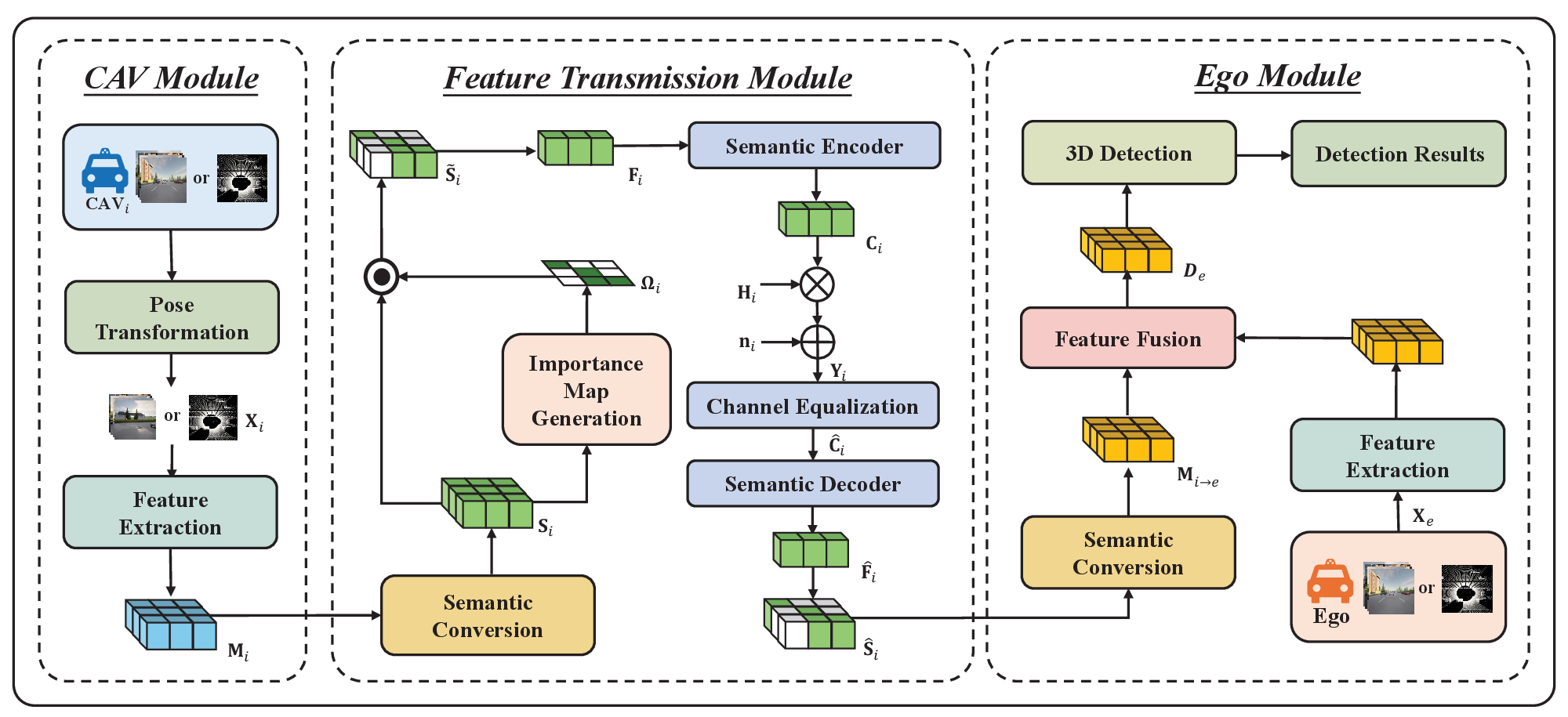}
  \caption{An illustration of the proposed CMSC framework for heterogeneous collaborative perception.}
  \label{fig2}
\end{figure*}

This section details the proposed CMSC framework for heterogeneous collaborative perception. As illustrated in Fig. 2, we consider a vehicle-to-vehicle (V2V) scenario where an ego vehicle receives perceptual features from multiple collaborating CAVs with two possible sensor modalities, i.e., camara and LiDAR. The overall architecture comprises three primary components: the CAV module, the feature transmission module, and the ego module.

\subsection{CAV Module}

In heterogeneous collaborative perception scenarios, there exists a set of sensor modalities $m \in \{l, c\}$, where $l$ denotes LiDAR and $c$ denotes multi-view RGB cameras. To process data from diverse sensor modalities, we employ a modality-specific feature extraction backbone, denoted as  $\mathbf{\Gamma}_{\theta_{m}}(\cdot)$. For LiDAR raw data, we utilized a network based on PointPillars~\cite{pointpillars}, parameterized by $\theta_l$. For camera raw data, we adopt a network based on Lift-Splat~\cite{lift}, parameterized by $\theta_{c}$. The raw perceptual data $\mathbf{X}_{i}$ of the $i$th CAV, which is spatially aligned with the ego vehicle's coordinate system, is then fed into its corresponding feature extraction network to extract bird’s-eye-view (BEV) feature maps $\mathbf{M}_{i}$, given by
\begin{equation}
    \label{eq:feature_extraction}
    \mathbf{M}_{i} = \mathbf{\Gamma}_{\theta_{m_i}}( \mathbf{X}_{i})\in \mathbb{R}^{H \times W \times C}, \quad \forall i \in \mathcal{I},
\end{equation}
where $m_i\in\{l,c\}$ denotes the sensor modality of the $i$th CAV,  $H$ and $W$ are the size of each feature map and $C$ denotes the number of channels. $\mathcal{I}$ is the set of collaborating CAVs.

\subsection{Feature Transmission Module}
The transmission pipeline begins by mapping the features $\mathbf{M}_{i}$ from their source modality $m_i$ into a standard semantic space $s$ using a semantic converter $\mathbf{\Gamma}_{\alpha_{m_i\rightarrow s}}(\cdot)$, parametrized by $\alpha_{m_i\rightarrow s}$. The parameters for this conversion depend on the source modality, with $\alpha_{l\rightarrow s}$ used for LiDAR features and $\alpha_{c\rightarrow s}$ for camera features. The conversion process is represented as
\begin{equation}
    \label{eq4}
    \mathbf{S}_{i} = \mathbf{\Gamma}_{\alpha_{m_i\rightarrow s}}(\mathbf{M}_{i}), \quad \forall i \in \mathcal{I},
\end{equation}
where $\mathbf{S}_{i} \in \mathbb{R}^{H \times W \times C}$ denotes the converted feature map in the standard semantic space.
Subsequently, a semantic importance-aware selector $\mathbf{\Gamma}_{\phi}(\cdot)$ retains only the most task-aware features from the feature map to reduce data volume to be delivered, represented by
\begin{equation}
    \label{eq5}
    \widetilde{\mathbf{S}}_i = \mathbf{\Gamma}_{\phi}(\mathbf{S}_i),\quad \forall i \in \mathcal{I},
\end{equation}
where $\widetilde{\mathbf{S}}_i \in \mathbb{R}^{H \times W \times C}$ retains the features in the selected spatial positions, with those in other positions being set to zero. These features are gathered into a dense tensor $\mathbf{F}_{i} \in \mathbb{R}^{K \times C}$, where $K$ denotes the number of selected spatial positions. The semantic encoder $\mathbf{\Gamma}_{\beta}(\cdot)$ then maps $\mathbf{F}_{i}$ to complex-valued symbols, given by
\begin{equation}
    \label{eq7}
    \mathbf{C}_{i} = \mathbf{\Gamma}_{\beta}(\mathbf{F}_{i}),\quad \forall i \in \mathcal{I},
\end{equation}
where $\mathbf{C}_{i} \in \mathbb{C}^{K \times C}$ denotes the encoded complex symbols. After wireless transmission, the received signal at the ego vehicle is modeled as
\begin{equation}
    \label{eq8}
    \mathbf{Y}_{i} = \mathbf{H}_i \odot \mathbf{C}_{i} + \mathbf{n}_i,\quad \forall i \in \mathcal{I},
\end{equation}
where $\mathbf{Y}_{i}\in \mathbb{C}^{K \times C}$ is the received signal, $\mathbf{H}_i \in \mathbb{C}^{K \times C}$ denotes the channel gain matrix, $\odot$ represents the Hadamard product, and $\mathbf{n}_i \in \mathbb{C}^{K \times C}$ is the additive white Gaussian noise (AWGN) with i.i.d. components distributed as $\mathcal{CN}(0, \sigma^2)$.

The received signal $\mathbf{Y}_i$ undergoes channel equalization to mitigate the effects of the channel's adverse effects, yielding the recovered symbols $\mathbf{\hat{C}}_i$. These symbols are then decoded by the semantic decoder $\mathbf{\Gamma}_{\eta}(\cdot)$, represented as
\begin{equation}
    \label{eq9}
    \mathbf{\hat{F}}_i = \mathbf{\Gamma}_{\eta}(\mathbf{\hat{C}}_i),\quad \forall i \in \mathcal{I},
\end{equation}
producing the reconstructed features $\mathbf{\hat{F}}_i \in \mathbb{R}^{K \times C}$. Finally, $\hat{\mathbf{F}}_i$ is remapped to the original spatial dimensions using the position indices from feature selection, yielding $\hat{\mathbf{S}}_i \in \mathbb{R}^{H\times W\times C}$ where unselected positions are zero-padded.

\subsection{Ego Module}

The raw perceptual data acquired by the ego vehicle is denoted as $\mathbf{X}_{e}$, which is then processed by the feature extraction network $\mathbf{\Gamma}_{\theta_{m_e}}(\cdot)$ corresponding to the ego vehicle's sensor modality $m_e$, to derive the BEV feature map $\mathbf{M}_{e}$, given by
\begin{equation}
\label{eq:ego_extraction}
\mathbf{M}_{e} = \mathbf{\Gamma}_{\theta_{m_e}}(\mathbf{X}_{e}).
\end{equation}
On the other hand, the received standard feature map $\hat{\mathbf{S}}_i$ shared by the $i$th CAV is then converted to align with the ego vehicle's sensor modality through the semantic converter $\mathbf{\Gamma}_{\alpha_{s\rightarrow m_e}}(\cdot)$, represented as
\begin{equation}
    \label{eq12}
    \mathbf{M}_{i\rightarrow e} = \mathbf{\Gamma}_{\alpha_{s\rightarrow m_e}}(\hat{\mathbf{S}}_i),\quad \forall i \in \mathcal{I},
\end{equation}
where  $\mathbf{M}_{i\rightarrow e} \in \mathbb{R}^{H \times W \times C}$ forms the modality-aligned feature map. By fusing the feature maps $\mathbf{M}_{i\rightarrow e}$ and $\mathbf{M}_{e}$ using the fusion network  $\mathbf{\Gamma}_{\xi_{m_e}}(\cdot)$, we derive the fused feature map $\mathbf{D}_{e}$, given by
\begin{equation}
    \label{eq13}
    {\mathbf{D}_{e} = \mathbf{\Gamma}_{\xi_{m_e}}({\mathbf{M}}_{e}, \mathbf{M}_{i\rightarrow e})},\quad \forall i \in \mathcal{I},
\end{equation}
which consolidates both local and collaborative perceptual information. Finally, the modality-specific detection network $\mathbf{\Gamma}_{\delta_{m_e}}(\cdot)$ processes the fused features to produce the perception output $\mathbf{\hat{R}}$, which contains both classification and regression components, expressed as
\begin{equation}
    \label{eq:detection}
    \mathbf{\hat{R}} = \mathbf{\Gamma}_{\delta_{m_e}}(\mathbf{D}_{e}).
\end{equation}

\begin{figure*}[htbp]
  \centering
  \includegraphics[width=0.75\linewidth]{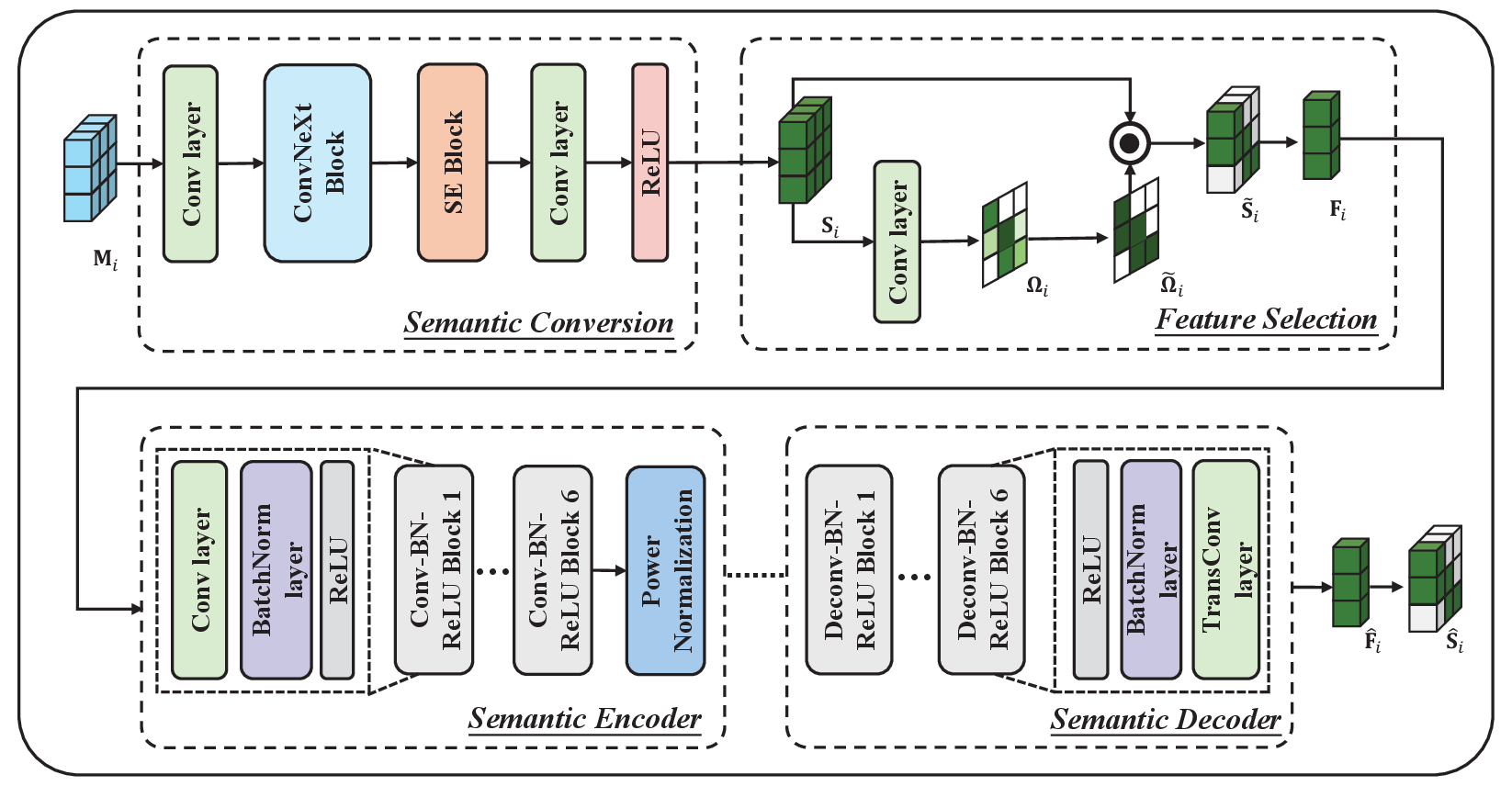}
  \caption{Network design of the feature sharing module.}
  \label{fig3}
\end{figure*}
\section{Cross-Modal Semantic Communication Design}
This section presents the architecture of the feature transmission module, shown in Fig.~\ref{fig3}, along with its training strategy and associated loss functions.
\subsection{Design of the Feature Transmission Module}

At the transmitter, the primary objective is to transform modality-specific features into a compact, semantically unified representation for efficient and robust transmission.

First, the semantic converter bridges the gap between heterogeneous sensor inputs. It employs a ConvNeXt~\cite{convnet} block for spatial feature extraction and a Squeeze-and-Excitation (SE)~\cite{se} block for channel-wise attention. This design dynamically refines feature channels to amplify perception task-relevant semantics while suppressing modality-specific noise, yielding a standardized feature map, $\mathbf{S}_i$ for the $i$th CAV.


Second, the importance-aware feature selection network enhances communication efficiency. This network processes the semantic feature map $\mathbf{S}_i$ using a convolution layer to generate a spatial importance map, $\mathbf{\Omega}_i$, which evaluates the task-relevance of features. Based on a predefined compression rate, this map is then binarized to create a hard-gating mask. Applying this mask selectively retains only the feature locations with the highest importance scores, resulting in a sparse feature map. This entire feature selection process is described by Eq.~\eqref{eq5}, which yields the sparse feature map $\widetilde{\mathbf{S}}_i$. The non-zero elements of $\widetilde{\mathbf{S}}_i$ are then gathered into a compact tensor $\mathbf{F}_i$, significantly reducing the data payload for transmission.

Finally, the semantic encoder resists channel impairments. Composed of cascaded Conv-BN-ReLU blocks, it compresses the feature vector $\mathbf{F}_i$ into complex-valued symbols. To comply with practical communication constraints, a power normalization layer is applied to the output, ensuring the transmitted symbols satisfy the channel's power budget.

At the receiver, the semantic decoder processes the received features from all CAVs. It mirrors the encoder's architecture, employing a series of Deconv-BN-ReLU blocks to reconstruct the feature vector.

\subsection{Loss Function and Training Strategy}
The proposed CMSC network is optimized using a three-stage training strategy.

In the first stage, we train the semantic converters while freezing other network parameters. For each modality $m\in \{l,c\}$, there exists a pair of semantic converters between $m$ and the standard semantic space $s$, $\mathbf{\Gamma}_{\alpha_{m\rightarrow s}}(\cdot)$ and $\mathbf{\Gamma}_{\alpha_{s\rightarrow m}}(\cdot)$. 
For a given training iteration focused on modality $m$, all vehicles in the scene, including the ego vehicle and all CAVs, are equipped with sensors of modality $m$, creating a homogeneous environment to learn a consistent mapping between $m$ and $s$. We employ a cycle-consistency framework that maps features across modalities and reconstructs them back to ensure consistency. The unidirectional conversions are defined as
\begin{equation}
    \mathbf{M}_{m \rightarrow s} = \mathbf{\Gamma}_{\alpha_{m\rightarrow s}}(\mathbf{M}_m), \quad \mathbf{M}_{s\rightarrow m} = \mathbf{\Gamma}_{\alpha_{s\rightarrow m}}(\mathbf{M}_s),
\end{equation}
where $\mathbf{M}_{m}$ and $\mathbf{M}_{s}$ denote the feature maps from modality $m$ and the standard space $s$, respectively. We define the standard space $s$ based on LiDAR modality and thus $\mathbf{M}_{s}$ is obtained by processing the raw point cloud data through the feature extraction network $\mathbf{\Gamma}_{\theta_l}(\cdot)$. Consequently, $\mathbf{M}_{m \rightarrow s}$ and $\mathbf{M}_{s \rightarrow m}$ are the corresponding unidirectionally converted features. 
The cyclical conversion is given by
\begin{equation}
    \mathbf{M}_{m \rightarrow s\rightarrow m} =  \mathbf{\Gamma}_{\alpha_{s\rightarrow m}}(\mathbf{\Gamma}_{\alpha_{m\rightarrow s}}(\mathbf{M}_m)),
\end{equation}
where $\mathbf{M}_{m \rightarrow s\rightarrow m}$ represents the cyclically converted features. 
Concurrently, the converted feature maps from all vehicles, $\mathbf{M}_{m\rightarrow s}$, are fused and processed by the detection network to produce a perception output, $\hat{\mathbf{R}}_s$.
The total loss for this stage integrates a perception loss with three consistency losses, formulated as
\begin{equation}
    \label{eq:stage1_loss}
    \begin{split}
        &L_{\text{stage1}}
        = L_{\text{cls}}(\hat{\mathbf{R}}_{s}, \mathbf{R})
           + \eta  L_{\text{reg}}(\hat{\mathbf{R}}_{s}, \mathbf{R})
         + \alpha L_{\text{mse}}( \mathbf{M}_{s},  \mathbf{M}_{m \rightarrow s})\\[4pt]
            &+ \beta L_{\text{mse}}( \mathbf{M}_{m}, \mathbf{M}_{s \rightarrow m})
            + \gamma L_{\text{mse}}(\mathbf{M}_{m}, \mathbf{M}_{m \rightarrow s\rightarrow m}),
    \end{split}
\end{equation}
where $L_{\text{cls}}$, $L_{\text{reg}}$, and $L_{\text{mse}}$ denote the classification loss, regression loss, and mean squared error loss, respectively. $\mathbf{R}$ represents the ground truth for 3D object detection. The hyperparameters $\eta, \alpha, \beta, \gamma$ balance the loss components.

In the second stage, the semantic converters are frozen, and we train the feature selection $\mathbf{\Gamma}_{\phi}(\cdot)$ and semantic codec modules, $\mathbf{\Gamma}_{\beta}(\cdot)$ and $\mathbf{\Gamma}_{\eta}(\cdot)$.
This training stage is conducted in the heterogeneous collaborative perception scenario as described in section II, with transmissions simulated over a Rayleigh fading channel. The objective is to learn an efficient and robust feature representation for transmission. The loss function balances perceptual accuracy with feature reconstruction fidelity and is defined as
\begin{equation}
    \label{eq:stage2_loss}
    L_{\text{stage2}} = L_{\text{cls}}(\hat{\mathbf{R}}, \mathbf{R})
           + \eta L_{\text{reg}}(\hat{\mathbf{R}}, \mathbf{R})
        + \mu L_{\text{mse}}(\hat{\mathbf{F}}_i, \mathbf{F}_i),
\end{equation}
where $\mathbf{F}_i$ represents the selected features before the semantic encoder, $\hat{\mathbf{F}}_i$ represents the reconstructed features after the semantic decoder, $\hat{\mathbf{R}}$ is the output from the ego vehicle's detection head, and $\mu$ is a balancing hyperparameter.

Finally, the third stage involves a global optimization of the entire framework. All parameters are unfrozen and fine-tuned in an end-to-end manner using the loss function $L_{\text{stage2}}$.

\section{Simulation Results}
We evaluate the performance of the proposed CMSC framework on the OPV2V-H dataset~\cite{heal}. The experimental setup comprises an ego vehicle engaged in collaborative perception with a fleet of CAVs equipped with diverse sensors. The evaluation metric is average precision (AP) for 3D object detection, calculated at intersection over union (IoU) thresholds of 0.5 and 0.7. Here, AP values indicates the overall detection quality by jointly considering precision and recall. Thus, higher AP values correspond to better detection performance.

Three baseline methods are introduced for comparison. The first two, referred to as Baseline-16QAM and Baseline-256QAM, represent existing heterogeneous collaborative perception methods without optimizing the communication process. They utilize the same feature selection module from our framework but employ a traditional communication pipeline consisting of uniform 8-bit quantization, 1/2-rate LDPC encoding, and 256-QAM or 16-QAM modulation. Since this pipeline is not end-to-end optimized for the perception task, it is inherently less efficient at preserving task-aware information under a constrained channel use budget. To specifically validate the necessity of the semantic conversion, we introduce a third baseline termed Baseline-JSCC. This method bypasses the semantic converters and directly applies a semantic codec to the heterogeneous features.
To ensure a fair comparison, the compression ratio $\lambda = \frac{K}{H \times W}$ is adjusted such that all methods maintain an equivalent number of channel uses, defined as
\begin{equation}
\label{eq:channel_uses}
{\text{Number of Channel Uses} = \frac{S_m \times \lambda \times 8}{R_c \times \log_2 M_c}},
\end{equation}
where $S_m$ is the total number of elements in the feature map, $R_c$ is the rate of the LDPC code, and $M_c$ is the modulation order. Additionally, we also establish a performance upper bound by considering an idealized scenario, where the complete feature maps from all CAVs are transmitted without any feature selection or compression successfully in their entirety.
\subsection{Performance analysis}
To demonstrate the necessities of cross-modal collaborative perception, we first evaluate the performance of our framework over a 20 dB AWGN channel across four distinct sensor modality configurations, as presented in Table~\ref{tab:hetero_results}. The compression ratio $\lambda$ is set to 0.06.
\begin{table}[t!]
\renewcommand{\arraystretch}{1.3}
\caption{Performance Evaluation Under Different Sensor Configurations.}
\label{tab:hetero_results}
\centering
\small 
\begin{tabular}{@{}cccc@{}}
\hline
\textbf{Ego Sensor} & \textbf{CAV Sensor} & \textbf{AP@0.5} & \textbf{AP@0.7} \\
\hline
LiDAR              & LiDAR                         & 0.961            & 0.904            \\
LiDAR              & Camera                        & 0.946            & 0.866            \\
Camera             & Camera                        & 0.586            & 0.300            \\
Camera             & LiDAR                         & 0.796            & 0.530            \\
\hline
\end{tabular}
\end{table}

\begin{figure*}[t]
  \centering
  \begin{minipage}[b]{0.49\textwidth}
    \centering
    \includegraphics[width=0.9\linewidth, height=0.52\linewidth]{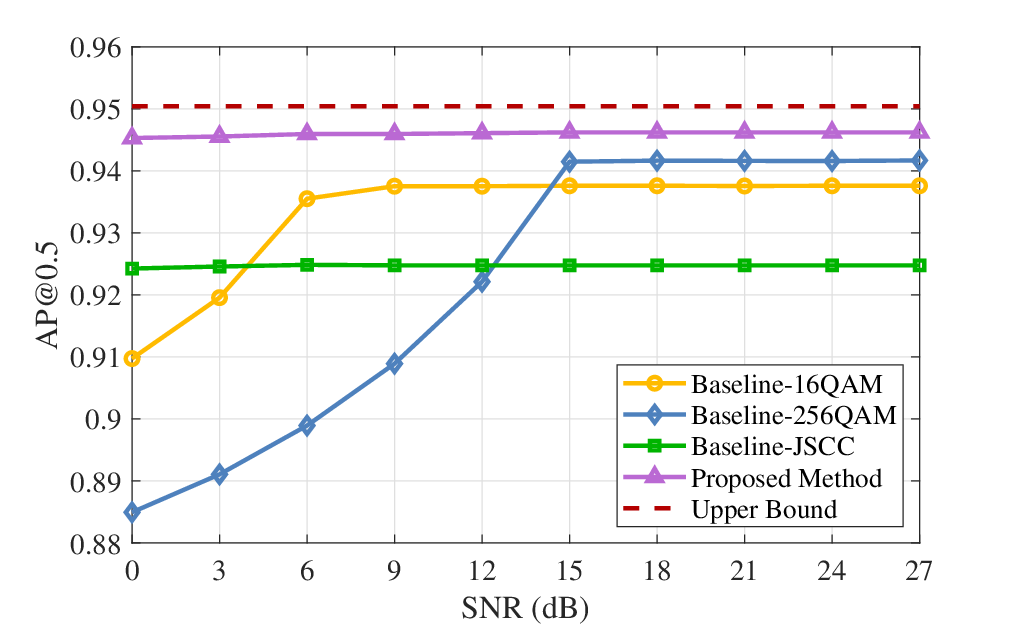}
    \centerline{(a) Ego vehicle equipped with LiDAR, AP@0.5}
  \end{minipage}
  \hfill
  \begin{minipage}[b]{0.49\textwidth}
    \centering
    \includegraphics[width=0.9\linewidth, height=0.52\linewidth]{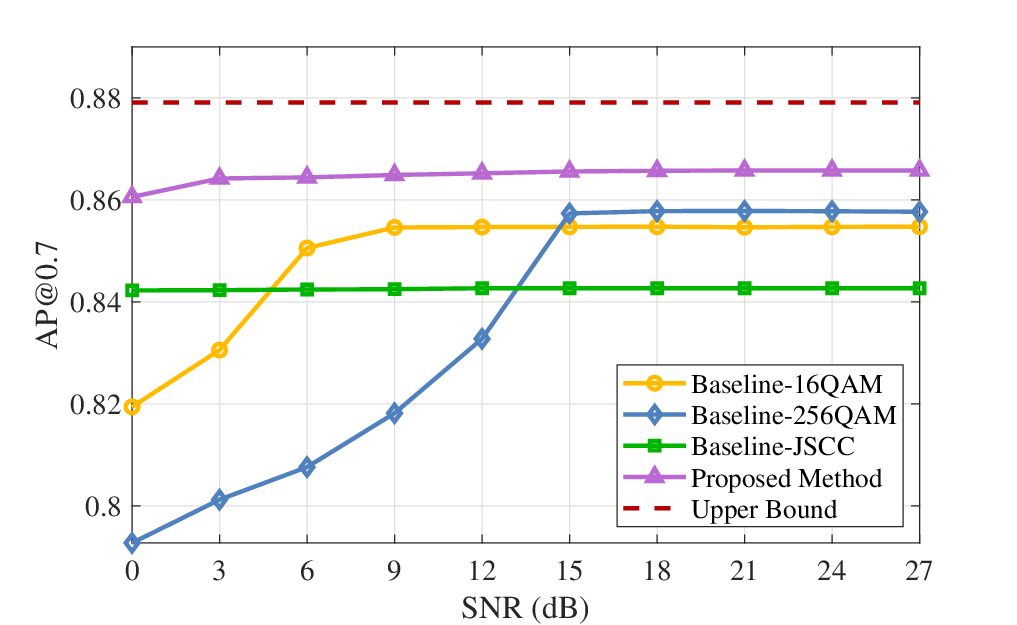}
    \centerline{(b) Ego vehicle equipped with LiDAR, AP@0.7}
  \end{minipage}
  
  \vspace{0.1cm}
  
  \begin{minipage}[b]{0.49\textwidth}
    \centering
    \includegraphics[width=0.9\linewidth, height=0.52\linewidth]{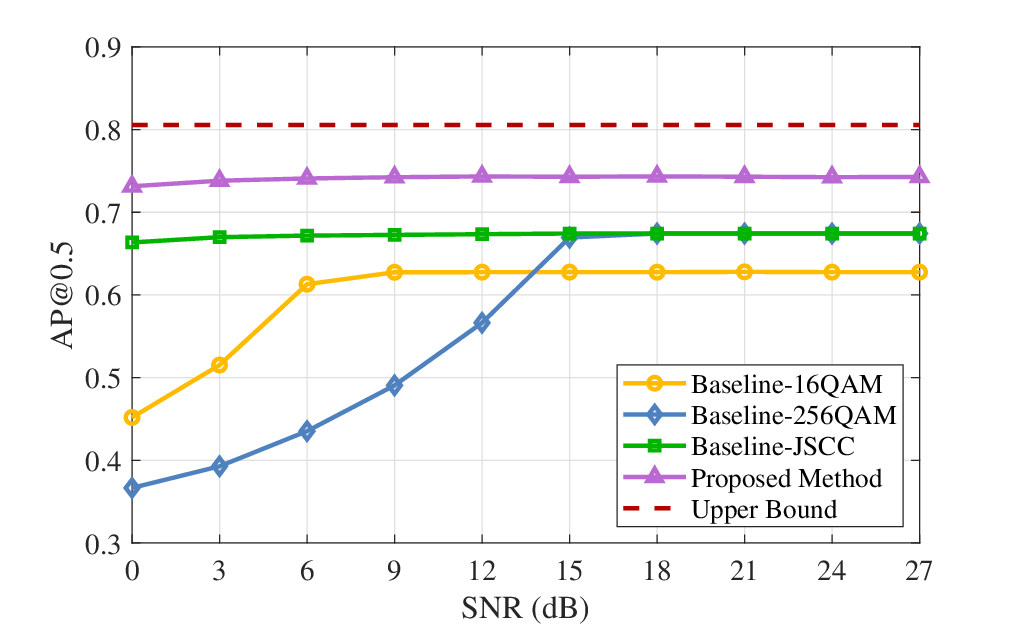}
    \centerline{(c) Ego vehicle equipped with Cameras, AP@0.5}
  \end{minipage}
  \hfill
  \begin{minipage}[b]{0.49\textwidth}
    \centering
    \includegraphics[width=0.9\linewidth, height=0.52\linewidth]{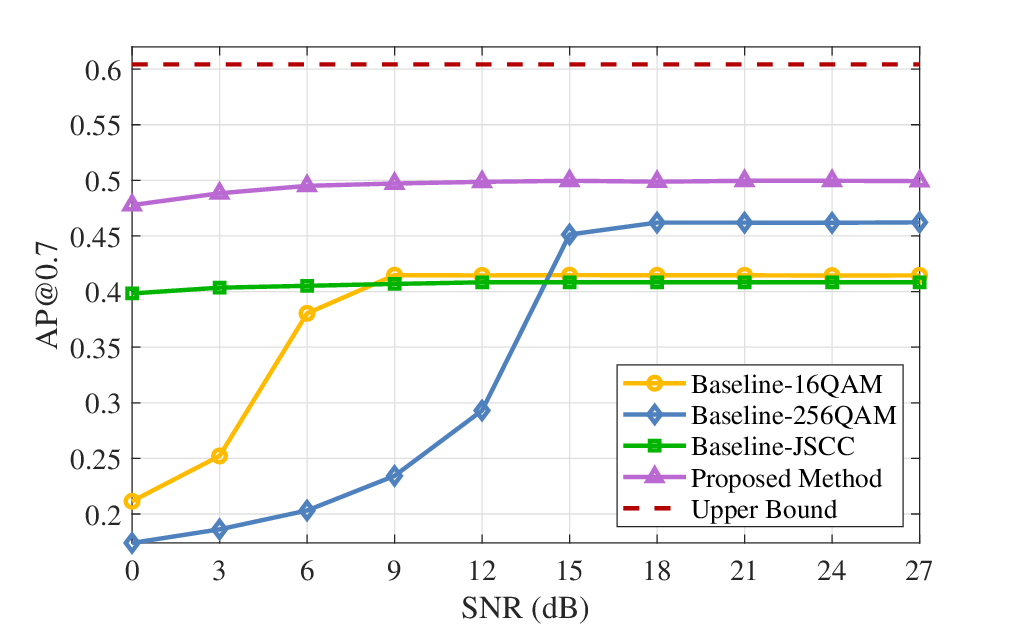}
    \centerline{(d) Ego vehicle equipped with Cameras, AP@0.7}
  \end{minipage}
  
  \caption{Performance comparison of our proposed method against baseline schemes and the upper bound at different SNR levels over an AWGN channel.}
  \label{AWGN}
\end{figure*}
As anticipated, the homogeneous LiDAR collaboration achieves the best performance due to the high precision of LiDAR sensors. Conversely, the homogeneous camera setup yields the lowest AP, reflecting the inherent limitations of cameras in 3D object detection. While all-LiDAR configuration represents best performance, it is unrealistic for large-scale deployment due to cost constraints. When a camera-equipped ego vehicle collaborates with LiDAR-equipped CAVs, its detection performance experiences a dramatic uplift. This result underscores the profound value of heterogeneous collaboration, as it allows the ego vehicle to effectively compensate for its weaker sensing modality by leveraging the high-quality perceptual information from its collaborators. This highlights the critical importance of the cross-modal collaborative perception enabled by our CMSC framework.


Subsequently, we illustrate the advantages of our CMSC framework by comparing its performance against the baseline methods across a range of SNRs over an AWGN channel.
Fig.~\ref{AWGN}(a) and (b) present the performance comparison over an AWGN channel when the ego vehicle utilizes a LiDAR sensor while the sensor modalities of CAVs are randomized. The compression ratio $\lambda$ for our method and Baseline-JSCC is set to 0.06, while it is set to 0.03 and 0.015 for Baseline-256QAM and Baseline-16QAM respectively according to Eq.~\eqref{eq:channel_uses}.As expected, the upper bound, which assumes ideal transmission of the complete feature map, achieves the highest perception performance.
In low-SNR regimes, the traditional baselines suffer from significant performance degradation. They demonstrate a pronounced cliff effect, which is highly sensitive to channel noise. Specifically, while the 16-QAM scheme is more robust at lower SNRs, it is surpassed by the higher-throughput 256-QAM scheme after 12 dB.
The Baseline-JSCC, benefitting from its end-to-end design, successfully mitigates the cliff effect in the very low-SNR region. However, its performance is fundamentally crippled by the lack of semantic conversion, which severely impairs the effectiveness of its semantic codec. This impairment is so pronounced that its performance is even surpassed by the traditional schemes at higher SNRs.
In contrast, our proposed CMSC method demonstrates consistent performance gains across all SNRs. The persistent advantage at high SNRs indicates that our CMSC network not only offers robustness against channel impairments but also learns a more efficient feature representation, better preserving critical semantic features.

Fig.~\ref{AWGN}(c) and (d) illustrate the results when the ego vehicle is equipped with RGB cameras while the sensor modalities of CAVs are randomized. A general performance degradation is observed across all methods compared to the previous scenario. This is attributable to the inherent lack of precise depth information in cameras, which is critical for 3D object detection.
In this configuration, the traditional baseline methods exhibit a more severe cliff effect. This vulnerability stems from the ego vehicle's weaker standalone perception capability, making it more reliant on the quality of features shared by collaborators and thus more sensitive to channel impairments. In this context, the channel robustness of Baseline-JSCC allows it to appear more competitive than in Fig.~\ref{AWGN}(a) and (b). The heightened reliance on collaborating information also explains the wider performance gap to the upper bound since a weaker ego vehicle has far more to gain from the ideal, lossless feature sharing represented by upper bound.
Despite the challenging conditions, our CMSC method maintains a consistent and substantial performance advantage across all SNRs. It can also be observed that the desirable properties of strong noise robustness and smooth performance transition are well preserved.

\begin{figure}[htbp]
  \centering
  \includegraphics[width=0.9\linewidth, height=0.52\linewidth]{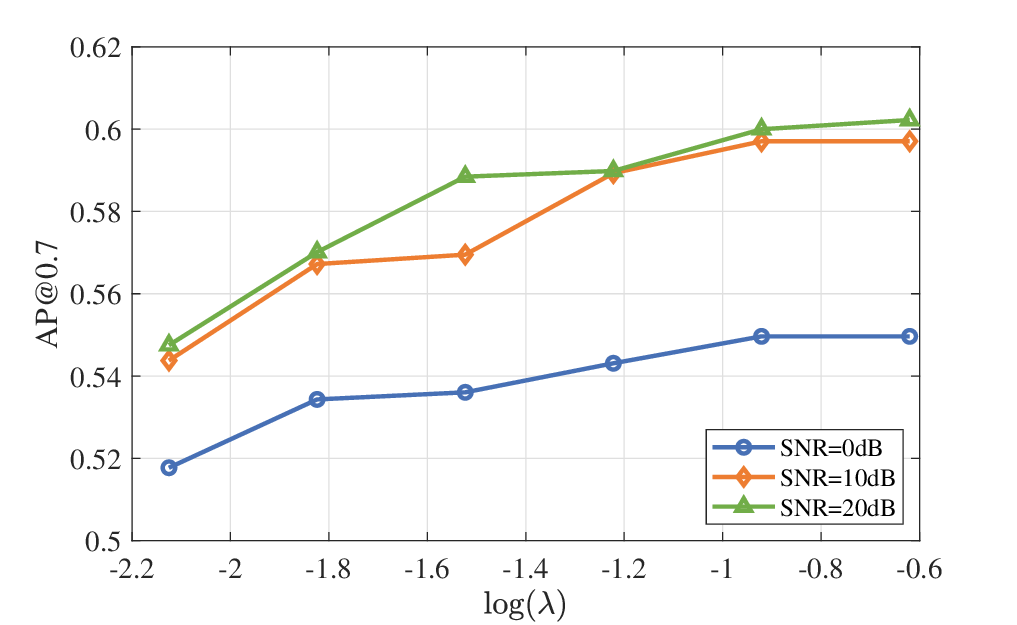}
  \caption{Performance of our proposed method compared with different compression ratios at different SNRs over a Rayleigh channel evaluated in terms of $\text{AP@}0.7$.}
  \label{cr}
\end{figure}

We further investigate the impact of the compression ratio $\lambda$ in terms of AP@0.7 over a Rayleigh fading channel. The ego vehicle's sensor modality was selected randomly for each instance. As depicted in Fig.~\ref{cr}, for any given SNR, the perception performance consistently improves as $\lambda$ increases because a larger $\lambda$ allows more semantic features to be transmitted. Moreover, higher SNR levels predictably result in better overall detection accuracy across all compression ratios.
Crucially, the performance curves begin to saturate beyond a certain compression ratio for each SNR level. This saturation point signifies that the majority of task-aware semantic information has already been effectively conveyed to the ego vehicle. Consequently, allocating more communication resources yields diminishing returns, as the additional transmitted data contributes negligibly to the perception task. This observed saturation behavior underscores the efficacy of the feature selection network. It demonstrates the network's ability to intelligently prioritize and transmit the most critical semantic features, achieving efficient performance even with constrained communication resources.

\section{Conclusion}	
This paper have introduced a novel CMSC framework to enable robust and efficient collaborative percpetion across diverse sensor modalities. The framework leverages a semantic converter to bridge the modality gap by mapping heterogeneous features into a unified, invariant semantic space. Furthermore, a semantic importance-aware JSCC strategy has been developed to selectively transmit only the most task-relevant information, ensuring both communication efficiency and resilience against channel impairments. Extensive experimental results demonstrate that our method significantly outperforms traditional schemes, particularly in low-SNR regimes where it effectively mitigates the cliff effect. 

\bibliographystyle{IEEEtran}
\bibliography{IEEEabrv,reference}

\vfill
\end{document}